Authentic Science Experiences with STEM Datasets: Post-secondary Results and Potential

Gender Influences

Andria C. Schwortz [a,b] and Andrea C. Burrows [a]

[a] University of Wyoming

[b] Quinsigamond Community College

Andria C. Schwortz, Physics & Astronomy, University of Wyoming / Natural Sciences,

Quinsigamond Community College, 670 W Boylston St, Box 224, Worcester, MA 01606,

aschwort@uwyo.edu, 1-508-854-7495, ORCID 0000-0003-1211-7620; Andrea C. Burrows,

Secondary Education, University of Wyoming, Laramie, WY, ORCID 0000-0001-5925-3596.

**Abstract**

**Background**: Dataset skills are used in STEM fields from healthcare work to astronomy research. Few fields explicitly teach students the skills to analyze datasets, and yet the increasing push for authentic science implies these skills should be taught.

**Purpose**: The overarching motivation is to understand learning of dataset skills within an astronomy context. Specifically, when participants work with a 200-entry Google Sheets dataset of astronomical data, what are they learning, how are they learning it, and who is doing the learning?

**Sample**: The authors studied a matched set of participants (n=87) consisting of 54 university undergraduate students (34 male, 18 female), and 33 science educators (16 male, 17 female).

**Design and methods**: Participants explored a three-phase dataset activity and were given an eight-question multiple-choice pre/post-test covering skills of analyzing datasets and astronomy content, with questions spanning Bloom's Taxonomy. Pre/post-test scores were compared and a t-test performed for subsamples by population.

**Results**: Participants exhibited learning of both dataset skills and astronomy content, indicating that dataset skills can be learned through this astronomy activity. Participants exhibited gains in both recall and synthesis questions, indicating learning is non-sequential. Female undergraduate students exhibited lower levels of learning than other populations.

**Conclusions**: Implications of the study include a stronger dataset focus in post-secondary STEM education and among science educators, and the need for further investigation into how instructors can ameliorate the challenges faced by female undergraduate students.

Processing large datasets is ubiquitous in science, mathematics, engineering, and technology (STEM) careers, and yet there is little to no formal teaching of necessary dataset skills at any level of STEM education. Not only are practicing scientists expected to know how to work with large amounts of data, but all people in modern societies face the use of datasets in their daily lives and careers. Without any formal training in the use of datasets, people must figure our how to use datasets "on the job" (Johri & Olds, 2014).

## Literature Review

This work intends to fill the gap in the literature that science education researchers understand conceptual learning but "have rather less to say about how to shape instruction in order to help students come to terms with the scientific point of view" (Scott, Asoko, & Leach 2007, p. 51). While a number of models describe how learners acquire the knowledge and skills to transition from novices to experts, efforts need to be made to put these models into practice in teaching (Bransford, Brown, & Cocking, 2000). The following literature review scans the topics of data education, the role of astronomy in STEM education, teachers and other science communicators, conceptual learning, and the relevance of Bloom's Taxonomy to the study's three-phase activity.

**The Need for Dataset Skills**

In the modern age, researchers often find themselves with more data than they can analyze by hand. While relevant literature generally does not re-articulate nor define the term "dataset" (e.g., Abello, Pardalos, & Resende, 2002; Brunner, Djorgovski, Prince, & Szalay, 2002; Leskovec, Rajaraman, & Ullman, 2011), here we will use it to refer to any collection of data, often arranged in a spreadsheet with each column containing a descriptor and each row a case. The use of computers to analyze large datasets allows not only for faster analysis, but also for the discovery of trends and patterns that may not be apparent without the use of computers. Data mining is this process of sifting through large amounts of data, finding patterns, making predictions, and explaining underlying properties of the dataset (Whitten, Frank, & Hall, 2011).

Within the realm of astronomy, for example, automated surveys allow for the collection of large amounts of data, more than can be analyzed object-by-object. Astronomers therefore must analyze these datasets en masse by programming their own data analysis tools. Assisting in this endeavor, any dataset created in the USA with government funding becomes public, allowing astronomers to address questions that the original researchers may not have considered.

Because using datasets is inevitable in STEM settings, dataset skills are crucial to undergraduate students in STEM fields from astronomy to zoology. Even at the pre-collegiate level, USA students are asked to demonstrate skills consistent with dataset use (CCSS, National Governors Association Center for Best Practices, 2010; NGSS Lead States 2013). The challenge for researchers is locating when and where dataset skills teaching occurs in STEM fields.

Formal education in dataset skills is lacking, and as a result students are underprepared to use these skills. Students report rarely using computers in any capacity in a school context, and when they do it is typically teacher directed (Danaia, McKinnon, & Fitzgerald, 2017). Perhaps due to this rare usage of computers, students of all levels have difficulty performing data analysis ranging from straightforward tasks like graphing (Jackson, Edwards, & Berger 1993), to more nuanced analysis of data (Wallace, Kupperman, & Krajcik 2000). Entry-level tools for analyzing sets of data such as Google Sheets, although insufficient for research in STEM fields, allow the user to view the data in rows and columns to search for trends, and have the ability to quickly and easily produce assorted graphs. These graphs can either be visually inspected (for trends and outliers) or analyzed by built-in tools that can fit a function (e.g. linear, polynomial, exponential).

STEM instructors do not teach the skills necessary for analyzing large datasets in a systematic way, with few pre-collegiate or post-secondary schools and programs in the USA requiring any programing or data analysis courses. As a result, the transition from novice to expert during the pre-collegiate to post-secondary and beyond is a haphazard process, with each individual learning the requisite skills and content on their own time or through performing research under the direct guidance of an expert during an ongoing project. To investigate dataset skills and use, the authors designed an expert/novice study where a STEM dataset was manipulated by different groups to test their dataset knowledge. The goal was to gain insight into how post-secondary learners progress in the acquisition of dataset expertise.

**Data Education**

There exists a plethora of reseach into topics tangential to data education, but a dearth in the topic itself. Tang & Tsai (2016) identified 454 papers in the five-year period of 2008-2013 on the use of educational technology. There has also been substantial research on the development of computational thinking and programming skills (e.g., Bailis, Hellerstein, & Stonebraker, 2015; Brennan & Resnick, 2012; Weintrop, et al., 2015). However, spreadsheets and other data management tools explicitly do not fit into these categories (Krajcik & Mun, 2014). Studies on the use of spreadsheets in accounting (e.g., Marriott, 2004) or math (e.g., Abramovich & Sugden, 2005; Sharma, 2006) are far from the content domain of this work. Numerous papers on the use of spreadsheets in engineering come closer, but tend to focus on describing the activity itself without examining the learning of the participants (Oke, 2004; Yamani & Kharab, 2001).

There is another gap in the literature about how people learn to work with large quantities of data in STEM using data management tools such as spreadsheets, and a lack of studies into the design of learning activities to teach these skills. While students can learn these skills once they enter their professional careers, this typically takes a longer amount of time, which is a luxury people often do not have in the modern fast-paced world (Johri & Olds, 2014).

**Astronomy Education**

The authors utilized astronomy as the content area to study the acquisition of dataset skills for a number of reasons, but mainly since its popularity allows everyone with an internet connection access to large datasets. Astronomy is a field which serves to capture the interest of the general public, with even non-scientists being able to participate in research through amateur groups and citizen science projects (Bailey, 2011). Collegiate students can more readily take introductory astronomy courses than physics due to the lower math level required (Bailey, 2011), thus making the course more relevant to non-majors.

Astronomy holds an important place among STEM fields due to its inherently interdisciplinary nature, but is often overlooked in literature reviews of STEM. For example, see the lack of astronomy mentioned in *The Handbook of Research on Science Education* (Abell & Lederman, 2013). Studies that compare different fields within STEM typically overlook astronomy (Sax, Lehman, Barthelemy, & Lim, 2016). Astronomy education research is often grouped with either physics or geology education research, however post-secondary astronomy does not fit well with either due to factors including the similarities and differences of mathematical analysis versus conceptual ideas, laboratory experiences versus fieldwork, and general content and topics covered (Bailey, 2011; Barthelemy, Mccormick, & Henderson, 2016). Only rarely is astronomy treated as a separate field (Ivie, White, & Chu, 2016).

The astronomy context of this study is discussed further in Table 1. Astronomy education is increasingly being studied at the level of individual concepts (e.g., Türk & Kalkan, 2017), or using computers to perform scientific inquiry (e.g., Danaia, McKinnon, & Fitzgerald, 2017). However, the use of computers to manipulate large datasets may allow students to learn different skills than those developed by working with individual problems (Berge, 1990).

Research astronomy is increasingly moving away from a realm of studying individual images, spectra, or time-based information, and towards studying large data sets derived from computerized analysis of those raw data products (Brunner, Djorgovski, Prince, & Szalay, 2002).

Data skills and techniques needed by future generations of astronomers include classification (grouping similar objects), identification of outliers, and data visualization (Brunner, Djorgovski, Prince, & Szalay, 2002). The need for handling big STEM datasets is ubiquitous, and thus it is important to investigate the process by which individuals transition from novices to experts in this domain (Bransford, Brown, & Cocking, 2000).

**Science Communicators**

Neither students nor non-technical audiences receive all of their information about STEM from instructors. Many other forms of STEM information and science communicators exist as well, such as "journalists, public information officers, scientists themselves" (Treise & Weigold, 2002, p. 311). Science content and the culture of science are frequently communicated (and miscommunicated) through mass media fiction ranging from classic novels to contemporary films. While numerous papers explore both the science content and the culture of science as portrayed in media (e.g., Barriga, Shapiro, & Fernandez, 2010; Gross, 2013; Steinke, 2005), the science knowledge of the creators of the media themselves has not been studied in as much depth. Therefore, such a group of science fiction writers are included in this study.

**Gender in STEM Learning**

Since females' and males' scores are shown below, a brief description about female learning in STEM can put some findings into context. It is well-known that girls' interest in STEM begins to drop off in middle school (Microsoft, 2017), that boys are more likely to take advanced placement exams in STEM fields at the end of high school than are girls, and at the undergraduate level males are more likely to major in STEM fields than females (Hill, Corbett, & St Rose, 2010). Causes for this are complex, with contributing factors including stereotype threat, low-self assessment, peer pressure, unconscious bias in teachers and family members, lack of female role models, and reduced access to preparatory courses or skills training (Hill, Corbett, & St Rose, 2010; Microsoft, 2017; Nissen & Shemwell, 2016). Girls are often socialized to be drawn to "helping" professions, and they don't see all STEM fields as fitting this image (Hill, Corbett, & St Rose, 2010).

The plethora of studies addressing gender in STEM learning (e.g., Brickhouse, Lowery, & Schultz, 2000; Nissen & Shemwell, 2016; Nyhof-Young, 2000; Nyström, 2007) is indicative of the fact that no easy solution exists. While investigations into gender in physics learning have shown promising results (e.g., Nieminen, Savinainen, & Viiri, 2013), there have been fewer studies examining gender in astronomy learning. In the field of computer education, it has been found that males control computers for a larger percentage of the time than females in mixed-gender groups (Day, Stang, Holmes, Kumar, & Bonn, 2016). Because of the progress in some of these fields, there is a risk that gender issues in science education will see reduced research in the future. But the lack of clear answers shows that instead it is more crucial than ever that this be addressed at the level of science teacher educators (Scantlebury & Baker, 2013).

**Novice / Expert Characteristics**

The differences between novices and experts in many subfields of science have been studied extensively. Bransford, Brown, and Cocking (2000) described six key characteristics distinguishing experts and novices. A modified version of these characteristics is presented in Table 2, along with descriptions of each and examples in the context of this study. In the domain of astronomy specifically, experts are better able to identify the underlying physical concepts which control the daily and yearly motions of the heavens (Bryce & Blown, 2012).

   **Bloom's Taxonomy**

Bloom's Taxonomy (Bloom, 1994) holds a key position in this work. This article uses the version of Krathwohl (2010), in which there are six fundamental levels of cognitive processes, listed in order of cognitive demand, defined as remembering, understanding, applying, analyzing, evaluating, and creating. Throughout this paper, the term "recall" will refer to the lower cognitive load or novice levels of remembering and understanding, and "synthesis" to the higher cognitive load or expert levels of evaluating and creating.

## Research Questions

The purpose of this study is to investigate novice and expert traits in undergraduate students and science educators learning to manipulate datasets within astronomy. Thus this study is guided by the overarching question of whether a three-phase, 1.5-hour activity could impact participants' short-term learning of the use of datasets in astronomy – or colloquially, "are they learning?" However this question is simplistic, as students can learn something from nearly any activity. The overarching question is then further refined into "*what* are they learning," "*how* are they learning it," and "*who* is learning it?" The following three-part question was developed.

How does this three-phase, 1.5-hour astronomy dataset activity impact participants' short-term learning scores on:

1. Skills and content focused questions? ("*What* are they learning?")

2. Novice (recall) and expert (synthesis) leveled questions? ("*How* are they learning?")

3. Subsamples based on (a) gender and (b) undergraduate/science educator status? ("*Who* is learning?")

## Theoretical Framework

The authors believe that learning is social and constructivist in nature, which informed the design of the three-phase learning activity. Studies continue to show the value of STEM learners working in groups (e.g., Eymur & Geban, 2017; Kutnick, et al., 2017). Participants discovered and built their own meaning in the data by discussing their ideas within their groups. However, for this paper's quantitative analysis, with limited qualitative references, the methodology aims to view the data through the numerical results of the participants' work. An understanding of how the participants constructed their knowledge is an important future step, but the first phase of understanding participants' expert and novice use of datasets starts with examining participants' initial results of manipulating the datasets.

## Methodology and Methods

This was largely a quantitative study, examining participant responses to eight multiple choice questions on a three-phase, 1.5-hour activity through pre/post-tests on astronomy content and dataset skills. On the pre/post-tests, participants performed the three tasks of data manipulation, data analysis, and data visualization (Weintrop, et al., 2015).

### Participant Sample

A total of 87 individuals participated in the pre/post-test study during 2014 through 2015. Fifty-four participants were undergraduate students in introductory astronomy courses at a large research university and a junior college in another state. Thirty-three science educators included in-service K-12 teachers, pre-service teachers in a graduate education program, a graduate student in a physical science field, and science and science fiction writers. The authors recruited science educators from professional development workshops. See Table 3 for more details.

### Experimental Procedure and Methods

After obtaining IRB approval, the authors worked with subsets of the participants for 1.5 hours on two dates over the course of two weeks, with data collection from different groups of participants taking place from June 2014 through February 2015. On the first contact date, participants took the pre-test (see Electronic Supplemental Materials, ESM), approximately a

half hour in time. On the second contact date, participants completed a three-phase activity (see

ESM) and the post-test (same as the pre-test), which took an average of an hour.

Data reported in this study are primarily quantitative, obtained from eight multiple-choice

questions on both astronomy content and data skills used as a pre/post-test. Validation of the

multiple-choice instrument was investigated by two methods. Firstly, the instrument was

reviewed by four individuals in astronomy and physics (a post-doctoral researcher, a lab

technician, a graduate student, and an advanced undergraduate student), and the questions

modified based on their feedback. Secondly, the free response questions (not used in this study)

probed similar topics to the multiple choice questions but in further depth.

For context, additional qualitative data were obtained by observing participant interactions during the activities, from video recordings of 20 groups of participants, and one-on-one interviews conducted between one week and three months after the completion of the activities. The activity required students to perform tasks of classification, identifying outliers, and data visualization, within the context of understanding quasars. The data skills questions included definitions of terms related to tables, classifying individual data points, and use of visualization tools. The astronomy content questions included definitions of terms related to quasars, drawing conclusions from data, and understanding causes of processes. The activity and data are further described in Table 4, with full details available in the supplementary materials.

**Analysis**

Analysis of variance (one-way ANOVA) was conducted to determine participants'

improvement from pre-test to post-test, along with computations of pre- and post-test means (as

a whole and by subsample), matched normalized gains, and effect size. Participants'

performance on questions, by level in Bloom's Taxonomy, was compared to their performance by skills or content. The authors split the participants into subsamples by gender, and by undergraduate / science educator status, and educational histories compared. Qualitative data including recordings, field notes, and interviews were obtained and a preliminary analysis performed. The gender composition of video recorded groups was tallied. The authors conducted preliminary coding of themes.

**Testing Measures**

The authors calculated the means, standard deviations, standard error, and p-values for pre- and post-test scores for the entire sample and subsamples of male, female, undergraduates, and science educators. Effect size was calculated as Cohen's d (Cohen, 1977). Matched normalized gains were calculated for each participant as per Hake (1998):

$$g_i = \frac{Post_i - Pre_i}{100\% - Pre_i}$$

The calculation for participants with a pre-test score of 100% had to be handled differently to prevent a divide by zero error. If their post-test score was also 100%, they were assigned a gain of 0. Otherwise their gain was calculated as follows.

$$g_i = \frac{Post_i}{Pre_i} - 1$$

Gain and effect size are correlated, so both have the same positive/negative sign or both are 0, with positive signs in both indicating the second group (i.e., post-test) has a higher score.

Effect size and gains were calculated from pre-test to post-test for all participants, by subsamples of gender and undergraduate/science educator status, by question novice/expert level, and by skills vs. content. For the categories of novice/expert level and skills/content, the results are presented in cross-tabular format, counting how many individual participants had positive, negative, or zero change.

Gains for each question category (overall, novice/expert level, and skills/content), and for each subsample (overall, by gender, by status), were analyzed by two-way multivariate ANOVA to further highlight the significance of the above results.

**Education Levels of Undergraduates vs. Science Educators**

Based upon the demographics survey, the number of undergraduates and science educators majoring in STEM fields were counted, as was the number of science educators who had taught a STEM subject for at least one year. These data are shown below in Table 5.

**Supporting Qualitative Data**

Qualitative data were obtained from audio and video recordings of and field notes about participant conversations during the activities, and from audio recordings of one-on-one interviews conducted within three months after the completion of the activities. All recordings were transcribed, and a preliminary coding for themes was performed. Since this paper is focusing on the quantitative analysis, only select instances are presented, with the purpose of supporting the quantitative data and providing insight into the reasons behind the findings. Additionally, for the recorded groups, a count was made of the number of males and females in each group, and whether male or female participants controlled the computer mouse and keyboard during the activity. Genders of participants in videos were identified visually.

## Findings

Through this dataset activity, participants improved their pre/post-test scores on an eight-question multiple-choice astronomy content and dataset skills test. More than half of the participants showed improvement in the categories of recall, synthesis, skills, or content; more than a third improved in both recall and synthesis; and nearly half improved in both skills and content. This activity was effective at short-term improvement on multiple choice questions on astronomy content and dataset skills, at both novice and expert levels. Many participants reported previous exposure to simple dataset usage, such as budgeting in spreadsheet programs.

The following information is organized by research question, with supporting qualitative data presented to triangulate the results. The specific subsample of female undergraduates is then explored in more depth.

**"Are They Learning?" (Overarching Question)**

Pre/post-test scores, matched normalized gains, and effect size (Cohen's d) are shown in Table 6 for the participants as a whole and the subpopulations. Unsurprisingly, learning did occur overall, with the pre-test means increasing from 61% to a post-test mean of 80%, with $p<0.01$ and a sample size of N=87. The matched normalized gain scores were 0.350 for all participants, and the overall effect sizes were 0.96 for all participants.

**"What Are They Learning?" (Question 1: Impact on Learning Skills vs. Content)**

Data pointed towards simultaneous improvement in both dataset skills and astronomy content, however more participants improved on content questions than on skills questions. Table 7 shows participants' pre- and post-test scores, gains, and effect size, broken down by skills and content, and by subsample. There was a large improvement on the content questions with gains of 0.412 or an effect size of 1.01. Skills questions also had a large gain, though a lesser one than of the content questions, of 0.349 or effect size of 0.73.

Table 8 is a cross-tabulation, showing how many participants and what percentage of participants improved their score (columns and rows indicated with +), stayed the same (0), or did worse (-), from the pre- to post-test. The crosstab shows that 43 individuals or 49% of the participants improved in both categories of skills and content, 14 individuals or 16% exhibited no change in either category, and 11 individuals or 13% did worse on one or both of skills and content.

**"How Are They Learning?" (Question 2: Impact on Learning by Novice/Expert Level)**

Data pointed towards simultaneous improvement in both the novice/recall and expert/synthesis level questions, as shown in Table 9. Participants scored higher in the category of recall questions on both pre- and post-tests (58% and 82%, respectively), than they did on the synthesis questions (36% and 51% respectively). Their improvement in both categories was large as well as being somewhat closer to parity, with participants exhibiting recall gains of 0.452 and effect size of 1.14, and synthesis gains of 0.351 and effect size of 0.70.

Table 10 uses crosstabs to count participants who improved or did worse in the categories of novice (recall) and expert (synthesis) level questions. While the largest category of participants (30 individuals or 34%) exhibited improvement on both novice and expert questions, 13 individuals or 15% showed no change in either category of questions. Nineteen participants (22%) did worse on one or both categories of novice or expert questions.

The following qualitative data support the finding that participants learned both novice and expert material. All sub-populations discussed recall-level ideas during the activity, such as definition of terms. In some groups one member came in with a higher level of recall knowledge, and individual was more likely to define the term for the other group members. Other groups were more likely to debate the meaning of terms and come to a consensus. Although few groups discussed synthesis-level ideas, it was witnessed in the one-on-one interviews. For example, one participant, Sarah, discussed how the number of quasars should depend upon redshift due to factors of both time evolution and volume observed, a synthesis level of understanding.

**"Who Is Learning?" (Question 3: Impact on Learning by Subsample)**

All subsamples, other than female undergraduates, showed statistically significant learning on the overall pre/post-test ($p<0.01$). Significance levels from ANOVA are shown first, followed by further detail of each subsample: subsamples by gender, whether undergraduates and science educators are meaningful groupings, and lastly subsamples by undergraduate/science educator status are displayed.

ANOVA significance levels for improvement from pre- to post-test are shown in Table 11. These are split by gender, undergraduate/science educator status, and interaction of these two categories, and are shown for the pre/post-test overall, recall vs. synthesis, and skills vs. content. No statistically significant interaction effect was found. Gender had a statistically significant effect on the improvement overall, and on the dataset skills questions. Undergraduate/science educator status had a statistically significant effect on the astronomy content questions. Table 12 explores whether pre- and post-test scores are distinct between subsamples. Differences were found ($p<0.05$) between the pre-test scores of male undergraduates and male science educators, and between post-test scores of undergraduates and science educators of all gender subsamples.

Next, the results by gender are explored in more depth. Both males and females had a large overall effect size, with males at 1.06 and females at 0.96, as shown in Table 6. The gain of male participants' was 0.458 (comparable to Hake's 1998 finding for active learning situations), while female participants' was significantly lower at 0.274 (comparable to Hake's lecture-based classrooms). ANOVA confirmed that these differences were statistically significant (Table 11).

The count of individuals in STEM fields shown in Table 5 indicates that undergraduates and science educators should be treated as different sub-populations. STEM fields comprised 15% of the undergraduates' majors, while 59% of science educators either had majored in STEM fields or were currently teaching STEM subjects. This showcases these two population's distinct exposure to STEM courses, including STEM content and skills, and thus they will be treated as distinct subsamples throughout this paper.

Having established that undergraduates and science educators are distinct populations, their pre/post-tests can now be compared. The difference in pre-test scores between undergraduates and science educators overall was not statistically significant (see Table 12), however science educators' post-test scores were 13% higher than that of the undergraduates. The science educators also showed more learning than did the undergraduates overall (Table 6), of both skills and content questions (Table 7), and of both novice and expert level questions (Table 9).

Qualitative data is now presented to help triangulate the quantitative findings above. Both undergraduates and science educators showed evidence of males dominating the group dynamics. During the activity itself, undergraduates formed either mixed-gender or all-male groups, while science educators formed either mixed-gender or all-female groups. Of the 20 video recorded groups, nine were mixed-gender, nine were exclusively male, and two were exclusively female. In the nine mixed-gender groups, three had only males controlling the computer, three had males controlling the computer the majority of the time (with females only touching the mouse occasionally), two had only a female controlling the computer, and one was unable to be determined due to the video angle.

Conversations in mixed-gender groups tended to be dominated by the males. For example, one five-person group of science educators consisted of two males and three females. In this group, the two males and one female continually interrupted each other: when one of these three individuals suggested an idea, the other two talked over the speaker to have his/her opinions heard. The remaining two females, on the other hand, spoke quietly and only to each other. During the transcription of this group, it was noted that "the males routinely interrupt or speak over the females," and the softer volume of the females meant that they were not heard as clearly on the recording. In the first 10 minutes, there were five instances of females speaking inaudibly and to themselves. This was not unique to the science educators, as similar behaviors occurred in the undergraduate mixed-gender groups as well.

**Subsample Interactions: Female Undergraduates**

In addition to examining each subsample individually, the authors also examined the interactions of these subsamples. Two-way ANOVA indicated no additional interaction effect between gender and undergraduate/science educator status, indicating that gender does not have a significantly different effect on undergraduates than it does on science educators. However the data show that the effects of gender and of student/professional status do compound. The learning of female undergraduates lagged behind that of both their male undergraduate peers, and their female science educator counterparts. This held true both for their overall scores, and for the dataset skills questions in specific. Female undergraduates only bested the male undergraduates in their improvement on novice questions. There were no categories in which they learned more than the female science educators.

As shown in Table 6, female undergraduates were the only subsample that did not show statistically significant improvement on the overall pre/post-test (p=0.52). This group's effect size is moderate at 0.62, compared to others' effect size all being greater than 1. The female undergraduates' gains were drastically lower than all other groups at 0.084 compared to the other subsamples' gains over 0.4.

When examining the supporting qualitative data, it was found that only the undergraduates discussed stereotypically male and female behavior, or used gender-based insults and jokes. For example, one female undergraduate in a mixed-gender group referred to writing a lengthy description as "a female thing," with implications that females are better at secretarial tasks rather than scientific ones. In another undergraduate group composed of only males, participants repeatedly made both gendered and sexual reference, including "yo momma" jokes and sexual innuendos. No gender-based conversations were recorded or observed with the science educators.

## Discussion and Conclusions

Many STEM majors currently get their first introduction to dataset analysis by being involved in research projects where they pick up skills while learning. Today many students, especially STEM majors, face prospects of struggling to teach themselves these skills in isolation. Research advisors currently decide between throwing their research students into the data without assistance, and using their valuable time to teach each student these skills one-on-one. This pattern does not need to remain the norm as dataset skills can and should be taught systematically in the course of undergraduate classes, or as part of STEM lab experiences. As astronomy often serves as a gateway to other sciences both for students and for the general public, astronomy datasets, such as SDSS, are vital for classroom and general use.

This activity demonstrates the feasibility of formal education in STEM dataset skills, and that it is reasonable to teach both STEM content and dataset skills simultaneously. Astronomy content specifically can be used as a tool to teach dataset skills in a way that students are likely to find engaging and interesting.

This dataset activity allowed participants to learn both at the novice (recall) level and at the expert (synthesis) level, and with non-sequential learning dominating. Learning is not sequential, and teaching in STEM disciplines need not always follow a simple path from "easier" content to "harder." Participants showed evidence of learning both content and skills, providing a proof of concept that a single activity can address both regimes of STEM learning.

Participants self-identified has having little prior knowledge in STEM datasets, and few science educators had previous experience with astronomy content. The data reflects this lack of experience, but does not preclude the possibility of students coming into the classroom with varying levels of prior knowledge. Materials that contain multiple levels of content can provide differentiated opportunities that benefit students at all levels.

Despite the level at which science educators teach, they are not immune to the need for instruction. Even with higher levels of STEM education, many science educators knew no more at the start than their undergraduate counterparts. These motivations however may be a contributing factor to the science educators' greater learning than the undergraduate participants.

The success of the science educators is not mirrored with the entirety of the undergraduate participants. We will now further discuss the performance of undergraduates by gender. The numerical data of this study do not allow us to speculate as to causes of difference by gender, however preliminary analysis of qualitative data does shed some light on how society and educators may be failing to situate all students for learning.

Female undergraduates were the only population that did not demonstrate improvement on their overall test scores, or on the questions focused on expert level understanding or dataset skills. This cannot be attributed to small number statistics, as there were more female undergraduates than either female or male science educators. It also cannot be attributed to prior knowledge: female undergraduates' pre-test scores were comparable to that of the other participants, however their post-test scores remained essentially unchanged. Two-way ANOVA showed no interaction between the statuses of "female" and "undergraduate," indicating that while the effects of these two statuses may compound, there is no differing effect of gender upon undergraduates from that upon science educators.

Observations of the dynamic of female and male participants may shed light on the lack of learning by female undergraduates. In this study, when female undergraduates worked in mixed-gender groups, they were never observed to control the computer mouse or keyboard. While it could be expected that male undergraduates would control the computers for a larger percentage of time (see literature review, above), the total domination of the computer by males is important to note.

The two instances of a female being the only one to control the computer in a mixed-gender group were both the individual Sarah, discussed previously. Sarah was a female science educator who participated in the study twice. She held an advanced degree in the physical sciences, while her male and female partners both times all held STEM education Bachelor's degrees. This study's complete lack of female computer use in undergraduate mixed-gender groups could be a factor in the females' lower post-test scores and learning.

Gendered insults from peers and self-deprecation may have been another important factor in the male-female interactions in this activity. Peer comments about gender are a classic example of external factors such as hostile environments that discourage females in STEM. And the gendered self-deprecation fits the known pattern of females in STEM possessing a lower self-assessment of their abilities than do males. Only seven of the 54 undergraduates (13%), or three of the 18 female undergraduates (17%), were in courses where the primary instructor was a woman. In other words, only 17% of the female undergraduates were exposed to women role models in the highest position of authority in their courses. These external and internal factors might have worked together to discourage female learning in STEM.

While males dominated computer usage in both undergraduate and science educator mixed-gender groups, female science educators appear to have overcome this difficulty. Those females who become science educators have managed to survive the leaky pipeline, indicating that they have managed to compensate for the deficiencies in the educational system that placed them at a disadvantage. During the activity itself, female science educators did not have to contend with the toxic environment of gendered comments that the female undergraduates faced, freeing them to concentrate on the content. Male science educators' apparent greater maturity than their undergraduate counterparts, as none were observed making gender-based jokes or insults, also created a less hostile environment for female science educators. In future work, the authors will further examine the qualitative data to determine and explore the role of these gender dynamics in the participant learning.

**Limitations**

The study highlighted only short-term learning and retention in content and skills acquisition of participants. Validation of the multiple choice questions was unable to be confirmed via the free response questions, as unfortunately the written responses showed significant evidence of test fatigue, such as many pre-test questions being left blank, and post-test responses reading "same as on the pre-test."

While the total number of participants allowed for statistical analysis, the number of male undergraduates was approximately double that of any of the other subsamples of female undergraduates, male science educators, and female science educators. However, it is unlikely that the lesser of learning among the female undergraduates is due exclusively to small number statistics, as the groups of male and female science educators had comparable numbers and they did exhibit learning.

Preliminary analysis of the data began after all undergraduate data was collected, and approximately half of the science educator data. This preliminary analysis indicated that females were struggling to learn in mixed-gender groups. As a result, for subsequent iterations of the activity, science educator participants were assigned to single-sex groups to reduce any potential harm that might be caused to females in mixed-gender groups.

This study presents only preliminary qualitative data. A deeper understanding of motivations of individuals, and interactions between them, will be explored in a future work.

## Implications

The findings of this study imply the need for further dataset education, both at the introductory undergraduate level, and in professional development opportunities for pre-service teachers, in-service teachers, and other science communicators. If STEM fields require the use and analysis of datasets, then it is essential to educate various populations in these skills in addition to the content. K-12 teachers and students need access to and instruction about classroom activities. Science fiction writers, who can be responsible for inspiring young adults into science careers, should know the reality of scientific work to communicate it to their audiences. Undergraduates taking science electives should learn these skills in addition to the content. Students definitely need to learn these skills to compete in a 21$^{st}$ century workforce.

For the general public to learn either science content, or dataset skills, they need exposure to datasets. Astronomy dataset education is important for science educators, including pre-service and in-service teachers, as there are available, accurate datasets ready to use. It is reasonable to suggest that if teachers are to adequately teach astronomy content, they need to understand it at a level better than their students. Thus dataset professional development for teachers is needed.

The need for dataset education is even stronger among females than it is for males. Females are disproportionately leaving STEM fields, including astronomy, and they are not being well served in the classes they take. To better serve female STEM students of all levels, we need to find effective pedagogies and approaches to STEM education that do not disadvantage students based upon their gender. In order for students to be successful, whether in life or STEM careers specifically, they need not only a broad education in the sciences, but also to acquire the skills to work with datasets.

The fact that the science educators' gender-based behavior was so drastically different from that of the undergraduates is intriguing. Is this lack of such discourse among the science educators evidence that when the gender ratio is more balanced, gendered conversations naturally die out? Is this evidence that less mature STEM students are likely to influence who among their peers become STEM educators? Could increasing the numbers of female teachers in undergraduate STEM classes help balance an environment unwelcoming to female undergraduates? Regardless of the cause, it is incumbent on instructors to address gender-based behavior as part of classroom management as one step to level the playing field for female students.

Instructors of any gender can take a number of approaches to help female students overcome the barriers they face both inside and outside the classroom. If assigning groups, instructors can consider grouping by gender, or having two or more females per group. When learners work in mixed-gender groups, instructors can mandate rotation of computer control so all individuals have the opportunity to learn computer skills. Instructors should encourage professional language in the classroom, taking special care to stop jokes and insults. But self-derogatory talk by females and benevolent sexism should be discouraged as well. Role models and mentors should be provided for female students and students from other underrepresented groups to help these individuals see that they can overcome the stereotypes against them.

K-12 students in many disciplines – especially in STEM – could be exposed to large datasets and their content knowledge and skills would improve as evidenced in the findings of this study. Teachers are a key in creating and implementing real-world applicable lessons, and dataset instruction should be included for K-12 students for both exposure and future success in STEM field careers. Public scholarship (at any level) to educate diverse democracies includes access to all STEM field components. To be competitive in STEM and open options to all K-20 students, education on dataset analysis must become a standard part of any STEM curriculum. Lastly, we must focus on creating a "safe space" for everyone to learn about STEM datasets.

**Acknowledgements**

The authors would like to thank the participants of the study. The first author would like to thank the members of the University Wyoming Physics Department for their support during the course of this study.

This work was partially supported by the US Department of Education / Wyoming Department of Education under Grant #WY140202; by the US National Science Foundation Division of Undergraduate Education under Grant #1339853; and by the Space Telescope Science Institute under Grant #HST-EO-13237.001-A

Table 1

*Astronomy Context of the Study*

---

Quasars are a type of galaxy where the central supermassive black hole (SMBH) is actively accreting dust and gas from a surrounding disk. Quasar glow brightly due to friction of material circling the SMBH, and thus quasars can be seen from extreme distances. Radio light from quasars can yield information about the presence or absence of jets., while the redshift of visible light can determine distance. The quasars in the source sample have a mean distance from Earth of approximately 10.5 billion light years, expressed as redshift due to the expansion of the Universe and the Doppler effect. The distribution of quasars is important to understanding the properties of the Universe, as how tightly clustered they are has implications for their gravity and evolution over time. The location of each quasar (or any astronomical object) is recorded in spherical coordinates: two angular coordinates on the sky, plus distance.

The dataset used in this study is a subset of the Sloan Digital Sky Survey Data (SDSS) Release 5 (Schneider et al. 2007) catalog, plus data from the Very Large Array (VLA) Faint Images of the Radio Sky at Twenty-cm (FIRST) radio survey catalog. While the full catalog contains more than 30,000 quasars with more than 100 descriptors, the subset used with the participants contained 200 quasars (cases) in rows, and five descriptors in columns: quasar name, two angular coordinates, distance (given as redshift or $z$), and FIRST radio magnitude.

---

Table 2
*Expert Characteristics and Descriptions*

| Characteristic | General Description* | Example in Context** |
|---|---|---|
| 1. Recognize meaningful patterns | Patterns to numbers, data, images, etc. are interpreted for meaning, and are more easily remembered if they fit known meanings. | Creates appropriate graphs from raw data without prompting, and understands possible causes for the graph's shape, such as observational bias or properties of redshift. |
| 2. Organized knowledge | Address "big ideas" and methods in the field of knowledge. | Draw conclusions about quasar jets from radio brightness. |
| 3. Contextualized knowledge | Understands subtle meanings of the knowledge. | Distinguishes between lack of data about a galaxy's brightness, and a galaxy being faint. |
| 4. Knowledge retrieval | Knowledge is accessed quickly and effortlessly. | Tasks accomplished and questions answered quickly and correctly. |
| 5. Pedagogical content knowledge (PCK) and peer instruction | Ability to communicate knowledge to others and teach the material to them. | Group discussion pertinent and fruitful, with more knowledgeable partners easily explaining to less knowledgeable. |
| 6. Adaptability and metacognition | Applies knowledge to new situations, and can evaluate own knowledge and skills for gaps. | Apply knowledge about coordinates to mapping of quasars, or about redshift to distribution of quasars. |

Note. *Adapted from Bransford, Brown, & Cocking (2000). **The examples given are in the context of the activity performed in this study.

Table 3
*Number of Matched Participants by Category*

|                                | Male | Female | Total |
|--------------------------------|------|--------|-------|
| I. Undergraduates              | 34   | 18     | 54*   |
| A. Summer Astro course         | 5    | 2      | 7     |
| B. Summer online Astro course  | 4    | 3      | 7     |
| C. Fall Astro course           | 25   | 13     | 40*   |
| II. Science Educators          | 16   | 17     | 33    |
| A. PD 1                        | 6    | 2      | 8     |
| B. PD 2                        | 4    | 4      | 8     |
| C. PD 3                        | 3    | 7      | 10    |
| D. Science Writers             | 3    | 4      | 7     |
| Total                          | 50   | 35     | 87*   |

Note. *Two undergraduate participants' genders were undetermined.

Table 4
*Activity Summary*

| Task Name | Activity Description | Example Participant Tasks | Data Used in this Study |
|---|---|---|---|
| **Pre-Test And Post-Test** <br><br>(same instrument given at the beginning and end of the session) | 8 multiple choice questions:<br>• 3 recall<br>  ○ remembering and understanding, e.g., definitions of terms<br>• 2 application or analysis<br>  ○ Application or analysis, e.g., using information in data tables<br>• 3 synthesis<br>  ○ evaluation and creation, e.g., using evidence to support conclusions<br>• 3 skills, 4 content, & 1 combo questions | • Answer multiple choice questions<br>• Answer open response questions with words or sketches | • Correct # of multiple choice questions<br>• Correct # of recall vs. synthesis<br>• Correct # of skills vs. content questions |
| **Introduction** | • Instructor-led<br>• 10-15 minutes of background content: quasars and redshift.<br>• Presentation style similar to typical introductory astronomy lab. | • Listen<br>• Take notes | • N/A |
| **Participant Activity** | • Groups of 3-4 using 1 computer.<br>• Groups primarily self-selected, except for assigned final participant sets (based on preliminary data).<br>• Dataset composed of 200 rows of data in Google Sheets (Table 1).<br>• Each activity phase was completed (in order) before another phase was given out (to minimize looking up answers).<br>• Open-ended prompts to glean understanding<br>• Given increasing levels of instruction in the three-phase activity.<br>• Scaffolding was intended to approximate the steps taken by expert astronomers. | • Discussion<br>• Use sketches<br>• Count spreadsheet entries<br>• Highlight spreadsheet columns<br>• Sort spreadsheet columns<br>• Graph data<br>• Change axis titles<br>• Analyze graphs | • Recordings of 20 groups (all 3 phases)<br>• Field notes describing participant tasks, conversation topics |
| *Phase I* | • Open-ended questions asking the participants to characterize the data, and speculate on how it could be analyzed. | • Discussion<br>• Count spreadsheet entries | • N/A |
| *Phase II* | • Slightly more specific questions, such as how the data could be graphed, and what it would look like. | • Discussion<br>• Use sketches<br>• Control keyboard | • N/A |
| *Phase III* | • Precise directions to create a scatter plot of the quasars' positions in the sky and histograms of the quasars' redshifts and radio magnitudes. Detailed questions asking students to determine specific properties of the dataset from the graphs. | • Record answers<br>• Highlight spreadsheet columns<br>• Graph data<br>• Change axis titles<br>• Analyze graphs | • N/A |
| **Demographics Survey** | • Demographics information requested including race, nation of origin, gender, level of education, and field of study. | • Answer questions<br>• Control mouse and keyboard | • Demographics provided |
| **Interviews** | • One-on-one interviews, conducted one week to three months after the activity. All participants were invited to interviews, 9 undergraduates and 7 educators agreed and participated in interviews. | • Discuss experience with spreadsheets<br>• Reflect on responses in activity | • Identified themes |

Table 5

*Number of Participants by Category, and Number Majoring or Teaching STEM*

|  | Number in category* | Number in STEM* |
|---|---|---|
| I. Undergraduates | 71 | 11 (15%) |
| II. Science Educators | 37 | 22 (59%) |

Note. *Numbers include individuals with unmatched pre/post-tests, thus numbers are higher than in Table 3. Percentages are statistically significant at $p<0.01$.

Table 6

*Overall Pre/Post-test Scores, Standard Deviations, Gains, and Effect Size by Subsample*

| | **Pre-/Post-Test Scores** | | | **Gains / Effect Size** | | |
|---|---|---|---|---|---|---|
| | **Male** | **Female** | **Total** | **Male** | **Female** | **Total** |
| I. Undergraduates | 57±19 / 78±19* | 58±16 / 70±22$^{\dagger}$ | 58±18 / 75±20* | 0.467 / 1.12 | 0.084 / 0.62 | 0.285 / 0.86 |
| II. Science Educators | 73±20 / 91±10* | 57±27 / 85±13* | 65±25 / 88±11* | 0.438 / 1.18 | 0.475 / 1.37 | 0.457 / 1.22 |
| Total | 62±21 / 82±18* | 58±22 / 78±20* | 61±21 / 80±18* | 0.458 / 1.06 | 0.274 / 0.96 | 0.350 / 0.96 |

Note: *p<0.01, $^{\dagger}$p=0.052.

Table 7

*Pre/Post-Test Scores Gains, and Effect Size, by Skills/Content and Subsample*

| Skills | Pre-/Post-Test Scores | | | Gains / Effect Size | | |
|---|---|---|---|---|---|---|
| | **Male** | **Female** | **Total** | **Male** | **Female** | **Total** |
| I. Undergraduates | 63±20 / 81±25* | 68±22 / 69±25** | 66±21 / 76±26* | 0.515 / 0.79 | 0.088 / 0.06 | 0.326 / 0.44 |
| II. Science Educators | 73±21 / 84±15* | 59±22 / 79±20* | 66±22 / 82±18* | 0.344 / 0.61 | 0.426 / 1.02 | 0.386 / 0.80 |
| Total | 67±21 / 82±23* | 64±22 / 74±23* | 66±21 / 78±23* | 0.460 / 0.72 | 0.252 / 0.48 | 0.349 / 0.73 |
| **Content** | **Male** | **Female** | **Total** | **Male** | **Female** | **Total** |
| I. Undergraduates | 45±24 / 73±21* | 48±21 / 64±28* | 47±23 / 70±24* | 0.438 / 1.26 | 0.163 / 0.70 | 0.321 / 0.99 |
| II. Science Educators | 68±23 / 95±12* | 52±38 / 82±17* | 59±32 / 88±16* | 0.635 / 1.56 | 0.491 / 1.07 | 0.561 / 1.16 |
| Total | 52±26 / 80±21* | 50±30 / 73±25* | 52±27 / 77±23* | 0.501 / 1.19 | 0.322 / 0.86 | 0.412 / 1.01 |

Note. *p<0.05, **p=0.044.

Table 8

*Improvement on Skills vs. Content Questions*

| All (N=87) | | Skills | | |
|---|---|---|---|---|
| | | - | 0 | + |
| Content | - | 3 (3%) | 3 (3%) | 0 (0%) |
| | 0 | 2 (2%) | 14 (16%) | 4 (5%) |
| | + | 3 (3%) | 14 (16%) | 43 (49%) |

Table 9
*Pre/Post-Test Scores, Gains, and Effect Size, by Novice/Expert Level and Subsample*

| | **Pre-/Post-Test Scores** | | | **Gains / Effect Size** | | |
|---|---|---|---|---|---|---|
| **Novice/Recall** | **Male** | **Female** | **Total** | **Male** | **Female** | **Total** |
| I. Undergraduates | 57±28 / 78±29* | 54±26 / 81±23* | 57±27 / 78±28* | 0.412 / 0.76 | 0.481 / 1.16 | 0.407 / 0.80 |
| II. Science Educators | 69±28 / 83±17* | 53±24 / 92±15* | 61±27 / 88±16* | 0.302 / 0.64 | 0.735 / 2.05 | 0.525 / 1.24 |
| Total | 61±28 / 80±26* | 53±25 / 87±20* | 58±27 / 82±24* | 0.377 / 0.72 | 0.605 / 1.51 | 0.452 / 1.14 |
| **Expert/Synthesis** | **Male** | **Female** | **Total** | **Male** | **Female** | **Total** |
| I. Undergraduates | 33±18 / 46±22* | 31±21 / 43±25** | 33±19 / 45±23* | 0.324 / 0.64 | 0.250 / 0.49 | 0.287 / 0.56 |
| II. Science Educators | 46±24 / 63±11* | 33±29 / 59±19* | 39±27 / 61±15* | 0.438 / 0.92 | 0.471 / 1.08 | 0.455 / 0.98 |
| Total | 37±21 / 51±20* | 32±25 / 50±23* | 36±23 / 51±21* | 0.360 / 0.68 | 0.357 / 0.76 | 0.351 / 0.70 |

Note. *$p<0.01$, **$p=0.082$.

Table 10

*Improvement on Recall vs. Synthesis Questions*

| All (N=87) | | Novice/Recall | | |
| --- | --- | --- | --- | --- |
| | | - | 0 | + |
| Expert/Synthesis | - | 1 (1%) | 1 (1%) | 10 (11%) |
| | 0 | 4 (5%) | 13 (15%) | 13 (15%) |
| | + | 3 (3%) | 12 (14%) | 30 (34%) |

Table 11

*ANOVA significance levels for overall pre/post-test improvement, by recall/synthesis question level, and by skills/content questions – for all participants, by gender, by undergraduate/science educator status, and by interaction.*

| | | Significance ($p$) | |
| | | With interaction term | Without interaction term |
|---|---|---|---|
| | Overall | 0.007** | 0.004** |
| | Recall | 0.038* | 0.051 |
| Gender | Synthesis | 0.817 | 0.796 |
| | Skills | 0.022* | 0.012* |
| | Content | 0.097 | 0.074 |
| | Overall | 0.189 | 0.252 |
| | Recall | 0.586 | 0.693 |
| Status | Synthesis | 0.258 | 0.269 |
| | Skills | 0.532 | 0.679 |
| | Content | 0.041* | 0.045* |
| | Overall | 0.127 | - |
| | Recall | 0.172 | - |
| Gender *Status | Synthesis | 0.717 | - |
| | Skills | 0.060 | - |
| | Content | 0.607 | - |

Note. *$p<0.05$; **$p<0.01$.

Table 12

*Differences in pre-test scores and post-test scores between populations.*

| | | Pre-Test % Difference | Post-Test % Difference |
|---|---|---|---|
| **Male − Female** | All | 3.9 | 4.3 |
| | Undergraduates | -1.7 | 7.4 |
| | Science Educators | 15.3 | 5.3 |
| **Undergrads − Science Ed's** | All | 6.7 | -13.3* |
| | Male | -16.0* | -13.1* |
| | Female | 1.0 | -15.2* |

Note. *p<0.05.